\documentclass[12 pt]{article}
\usepackage{graphicx, amsmath, amssymb, cite, setspace, color}

\usepackage[top=1 in, bottom=1 in, left=1 in, right=1 in]{geometry}

\newcommand{\captionfonts}{\footnotesize} %make footnote font size smaller
\makeatletter 
\long\def\@makecaption#1#2{%
  \vskip\abovecaptionskip
  \sbox\@tempboxa{{\captionfonts #1: #2}}%
  \ifdim \wd\@tempboxa >\hsize
    {\captionfonts #1: #2\par}
  \else
    \hbox to\hsize{\hfil\box\@tempboxa\hfil}%
  \fi
  \vskip\belowcaptionskip}
\makeatother 

\newcommand{\T}{\tan T_{\text{co}}} 
\newcommand{\Ts}{\tan ^2 T_{\text{co}}} 

\begin{document}

%\preprint{PUPT-2287}

\title{Odds of observing the multiverse}
\author{A.~Dahlen\footnote{e-mail: adahlen@princeton.edu} \\ \\
\vspace{-.3 em}  \textit{\small{Joseph Henry Laboratories}}\\
\vspace{-.3 em} \textit {\small{Department of Physics, Princeton University}}\\
\textit {\small{Princeton, NJ 08544}}
}
\date{\today}

%\email{adahlen@princeton.edu}
%\affiliation{
%Joseph Henry Laboratories \\
%Department of Physics, Princeton University \\
%Princeton, NJ 08544}

\maketitle

\begin{abstract}
Eternal inflation predicts our observable universe lies within a bubble (or pocket universe) embedded in a volume of inflating space.  The interior of the bubble undergoes inflation and standard cosmology, while the bubble walls expand outward and collide with other neighboring bubbles. The collisions provide either an opportunity to make a direct observation of the multiverse or, if they produce unacceptable anisotropy, a threat to inflationary theory.  The probability of an observer in our bubble detecting the effects of collisions has an absolute upper bound set  by the odds of being in the part of our bubble that lies in the forward light-cone of a collision; in the case of collisions with bubbles of identical vacua, this bound is given by the bubble nucleation rate times ($H_{\rm{O}}/H_{\rm{I}})^2$, where $H_{\rm{O}}$ is the Hubble scale outside the bubbles and $H_{\rm{I}}$ is the scale of the second round of inflation that occurs inside our bubble.  Similar results were obtained by Freigovel \emph{et al.}~using a different method for the case of collisions with bubbles of much larger cosmological constant; here it is shown to hold in the case of collisions with identical bubbles as well.
\end{abstract}

%\pacs{04.00.00}

\section{Introduction}

In any eternal inflation scenario, bubbles (or pockets) of habitable universe form and expand in a background that continues to inflate.  In order to be habitable, the bubbles' interiors must undergo an additional, finite epoch of inflation before reheating.  In the meantime, the bubble walls continue to expand and eventually undergo collisions with other bubbles.  These collisions and their consequences, good and bad, are the focus of this paper.

\begin{figure}
\begin{center}
\includegraphics[width=130mm]{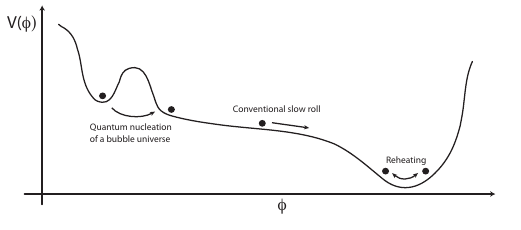}
\caption{An idealized potential $V(\phi)$ used as the model for eternal inflation in this paper.   Over some volume of space, a scalar field $\phi$ is initially trapped by a barrier in a false vacuum (left), causing the volume to inflate.  At an exponentially small rate, bubbles are formed inside of which the scalar field jumps to a value just to the other side of the barrier.  The field then slow-rolls down the potential, causing a round of inflation on the interior. Eventually, the field approaches a true minimum, oscillates around the minimum and heats up the bubble interior.  In the meantime, the bubble wall expands outwards into the eternally inflating space surround it and colliding with other bubbles.}
\label{SampPot}
\end{center}
\end{figure}

A simple realization of this is false-vacuum eternal inflation, in which scalar field trapped in the false vacuum of a potential like that in Fig.~\ref{SampPot} will undergo eternal inflation.  Through quantum tunneling, bubbles will nucleate, within which the scalar field has jumped to the other side of the barrier and has begun to roll towards the true minimum of the potential.  Initially, the spatial slices on the interior of the bubble are cold and open, so a second round of inflation within the bubbles' interiors is needed.  This second phase of slow-roll inflation, followed by reheating, ultimately produces a nearly flat, homogeneous, isotropic, hot big bang universe. At the same time, the bubble wall expands outwards, accelerating, and quickly reaches the speed of light.      

The same mechanism that nucleated one bubble creates an infinite number of other expanding bubbles, including an infinite number close enough to collide with the first.  The boundaries of these bubbles are massive and move ultra-relativistically, which suggests that the collisions are violent and can drastically disrupt the isototropy within our bubble.

The CMB is highly isotropic, so if eternal inflation produces unacceptably large anisotropy, it means that the theory is ruled out.  Alternatively, if the collisions produce a slight but still detectable effect, it provides us with the opportunity to confirm eternal inflation and the presence of a multiverse (of at least two vacua).  Further observations could then even verify the existence of a complex energy landscape, as suggested by string theory.  The two goals of this paper are to determine the likelihood of observing a sky as isotropic as ours in eternal inflation, and then, given that we live in the subset of observers for whom no effect has been observed thus far, to consider the likelihood that evidence for bubble collisions will be found as observations improve. 

These probabilities are evaluated by comparing the space-time volume on the interior of our bubble from which the collisions are visible to the volume from which they are not.  Of course, in so doing, the measure problem~\cite{Vilenkin:2006xv, Winitzki:2006rn, Aguirre:2006ak, Guth:2007ng} arises: namely, how does one choose the spatial slices and how does one deal with the fact that  both volumes are infinite?  By restricting the analysis is to a single bubble, there is a natural time-slicing, which addresses the first question.  Furthermore, although extracting a finite answer despite the infinities is not uncontroversial, a specific ratio can be constructed that is both finite and naturally suggested by the geometry.

This paper owes much to a series of papers beginning with Garriga, Guth, and Vilenkin (GGV)~\cite{Garriga:2006hw}.  They showed that, even though the interior of an isolated bubble is homogeneous and isotropic, this cosmological symmetry is broken by bubble collisions.  Specifically, they showed that the probability of observing a collision is not uniform over a spatial slice within a bubble, and that for observers who do see collisions, the distribution is anisotropic.  Furthermore, this breaking of translational symmetry is directly related to the motion of the observer with respect to the space-like surface on which inflation begins, a state about which little is known.  The fact that this surface must exist was proven previously~\cite{Borde:1993xh, Borde:2001nh}, but it was believed that any sensitivity to those initial conditions is erased if there is sufficient inflation.  However, the collision effect means that, contrary to expectations, memory of these initial conditions does not vanish with time, but instead persists in the form of a disruption of isotropy (which is how the effect earned its artistic name  ``the persistence of memory'').  A follow-up study, by Aguirre, Johnson, and Shomer~\cite{Aguirre:2007an} (AJS), calculated the angular distribution of collisions on our bubble wall for different observers within the bubble.  

A recent study by Freigovel \emph{et al.}~\cite{Freivogel:2009it} considered the case of collisions of bubbles of a cosmological constant that was close to that of the parent vacuum.  They found that the expected number of bubbles in our past light cone is proportional to the nucleation rate times the factor $(H_{\rm{O}}/H_{\rm{I}})^2$, where $H_{\rm{O}}$ is the Hubble scale outside the bubbles and $H_{\rm{I}}$ is the scale of the second round of inflation that occurs inside our bubble.  In this paper, it will be shown, using different techniques, that the same can be said for collisions with bubbles identical to our own, which is an important extension of the Freigovel \emph{et al.} result.  

There are three reasons why same-bubble collisions are worth considering.  First, collisions with our own type of bubble are guaranteed in a landscape, whereas the type of collision they consider do not necessarily occur.  Second, it is still not known which types of collisions will be most observable, so all types, and in particular same-bubble collisions, merit study.  Finally, there is an argument one can make that our parent vacuum should have an anomalously high tunneling rate to our type of vacuum.  It goes like this: consider a large number of drawers each containing differently colored socks, where some drawers have a large fraction of purple socks and others do not.  If you pick a random drawer and take a random sock and it is purple, then you likely chose a drawer with a large fraction of purple socks.  In this analogy, the different colors of sock represent different bubbles, and the drawers represent different parent vacua.  Of course, this argument assumes an even prior distribution over drawers, where as the correct prior to take over parent vacua is unknown.  However, it illustrates a selection bias that could make the nucleation rate to our type of vacuum larger than naively expected.

In their paper, Freigovel \emph{et al.} argue that the factor $(H_{\rm{O}}/H_{\rm{I}})^2$ can be large enough that bubble collision might be in our past light cone.  In that case, the question of whether they would be observable arises.  A key uncertainty is the observable remnants of a bubble collision.  What should the observer expect to see?  For better or worse, the answer is model dependent.  While collisions with other bubbles of our own vacuum are likely to produce matter, radiation, and gravity waves, collisions with bubbles that contain a different vacuum, could produce domain walls, or even more exotic features.  It seems likely that these new possibilities would make collisions with other vacua more observable than collisions with our own.

Several groups have considered the possible remnants of bubble collisions.  The earliest discussions of these collisions~\cite{Hawking:1982ga, Wu:1984eda}, allowed for no dissipation of energy at the collision site, which meant the bubble walls would pass through each other, then get pulled back and oscillate forever.  More recent models~\cite{Freivogel:2007fx, Chang:2007eq, Aguirre:2007wm, Chang:2008gj} have added energy dissiplation, but they have made the strong assumption that all the momentum is carried off from the collision by a narrow cone of radiation.  Within the confines of this assumption, Chang \emph{et al.}~\cite{Chang:2008gj} and Aguirre \emph{et al.}~\cite{Aguirre:2008wy} have shown that, indeed, collisions with identical bubbles are unlikely to be observed, whereas collisions with other types of vacua have several interesting observational features.  However, the issue is still not settled.

As a first step, the issue of collision remnants will be set aside.  After setting some basic notation and defining the geometry of the collisions in Sec.~2, the fraction of observers within our bubble who are in the forward light-cone of a collision---the maximal set of observers who could possibly detect a collision---is calculated in Secs.~3 and 4.  In order to do this, it will be assumed that the collisions do not significantly change the expansion rate within the bubble and that the number of observers is proportional to the spatial volume.  Up to the proposed resolution of measure issues, it is shown that, in the case of a potential like Fig.~\ref{SampPot}, for observers who have experienced more accelerated expansion than decelerated expansion since the bubble formed, this ratio is proportional to the nucleation rate, and therefore exponentially small.  Sec.~5  discusses different measures, and different observational signatures should the collisions be observable.

\begin{figure}
\begin{center}
\includegraphics[width=130mm]{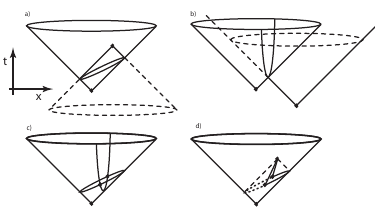}
\caption{A space-time diagram that represents the geometry of how an observer inside a bubble views a collision. a)  The light-cone drawn with solid lines represents a bubble whose walls are expanding outward at the speed of light.  Outside the bubble, the inflaton field is trapped in the false phase; inside, the field has settled into the true phase.  An observer in the interior looks back along his light-cone, which ultimately intersects the bubble wall and defines an ellipse.  b) The collision of two bubbles defines a hyperbolic scar on the bubble wall.  c)  If the ellipse and hyperbola intersect, they define the part of the observer's sky that is affected by the collision.  d)  Tracing the endpoints back, one can determine the angular scale of the collision.}
\label{CollGeo}
\end{center}
\end{figure}

\section{Geometry}

For the purpose of this calculation the fact that the bubble has finite radius at nucleation and that it takes some time for the wall to accelerate to relativistic speeds is ignored.  Instead, the bubble universe is treated as a light-cone emanating from a point-like nucleation site.  When an observer in the interior of this bubble looks back along his past light cone, as in Fig.~\ref{CollGeo}a, these two cones intersect and, in general, form an ellipse (or ellipsoid in the language of 3+1 dimensions). This ellipse represents the time slice which the observer would call the big bang.

Sometimes two bubbles will collide.  In this case the geometry is that of the two intersecting cones in Fig.~\ref{CollGeo}b.  The collision leaves a scar on our bubble wall, where the two cones intersect; this is the hyperbola in the figure.  Our observer will be able to detect the collision if the ellipse and the hyperbola intersect each other.  In that case, they define the section of our observers sky that is affected by the collision, as in Fig.~\ref{CollGeo}c.  Tracing the intersection points back to our observer, the angular scale of the scar on the sky can be determined, as in Fig.~\ref{CollGeo}d.  This angle was calculated by Aguirre, Johnson, and Shomer (AJS)~\cite{Aguirre:2007an}.

\begin{figure}
\begin{center}
\includegraphics[width=130mm]{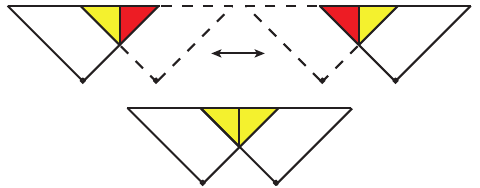}
\caption{A single collision divides our bubble into three regions, colored white, yellow, and red.  White observers are outside the light cone of the collision and, therefore, see nothing.  Yellow observers (light gray, in print) see a collision that occupies less than half their sky, and red observers (dark gray, in print) see a collision that occupies more than half.  Because the two bubbles are overlapping, the assignment of the colors yellow and red is not unique; an observer in the other bubble reverses them.  An appropriate interpretation is that the red volume has been excised by the collision.  The two bubbles are stitched together along the seam, and only white and yellow remain.}
\label{ExcRed}
\end{center}
\end{figure}

Viewing Fig.~\ref{CollGeo} from edge on, it looks something like Fig.~\ref{ExcRed}.  Collision with another bubble has divided our bubble into three different sections, which have been colored white, yellow, and red.  White observers are outside of the forward light-cone of the collision; they are ignorant of it and therefore see nothing.  Yellow observers see the scar, but are to the left of the hyperbola, which translates to the fact that the scar occupies an angular scale smaller than $\pi$; less than half of their sky is collision.  Red observers, on the other side of the hyperbola, are swallowed by the scar---it covers more than half of their sky.  

When two bubbles collide, they overlap, so the assignment in terms of yellow and red is not unique.  An observer in the other bubble just switches `yellow' with `red', as in Fig.~\ref{ExcRed}.  This overlap implies that an appropriate interpretation is that the collision has cut off the red region, and the two bubbles should be stitched together along the seam, with only yellow remaining.  A semi-infinite volume that would have been part of our bubble is excised by the collision.  This unambiguously accounts for the overlap and avoids any double-counting of volumes when an average is taken over all like bubbles.  To obtain an upper bound on the probability of observing a collision, the correct quantity to compare is the volume of yellow to the volume of white within our bubble.

It should be noted that the argument about excising red given above relies on the fact that the colliding bubbles are identical, and Earth could equally likely be in either.  However, in the case of collisions with different types of bubbles, a similar argument about excising a semi-infinite volume of our bubble can be made.  Such collisions produce domain walls, which as they move through the universe, either prevent the formation of structure or demolish any life that could have existed there.  A recent paper by Kleban \emph{et al.}~uses the motion of the domain wall to divide our bubble into inhospitable red regions, yellow regions from which the collisions are visible, and white regions which are outside of causal contact from the collision.  Likewise, they discount the semi-infinite red volume, and compare yellow to white.

\begin{figure}
\begin{center}
\includegraphics[width=100mm]{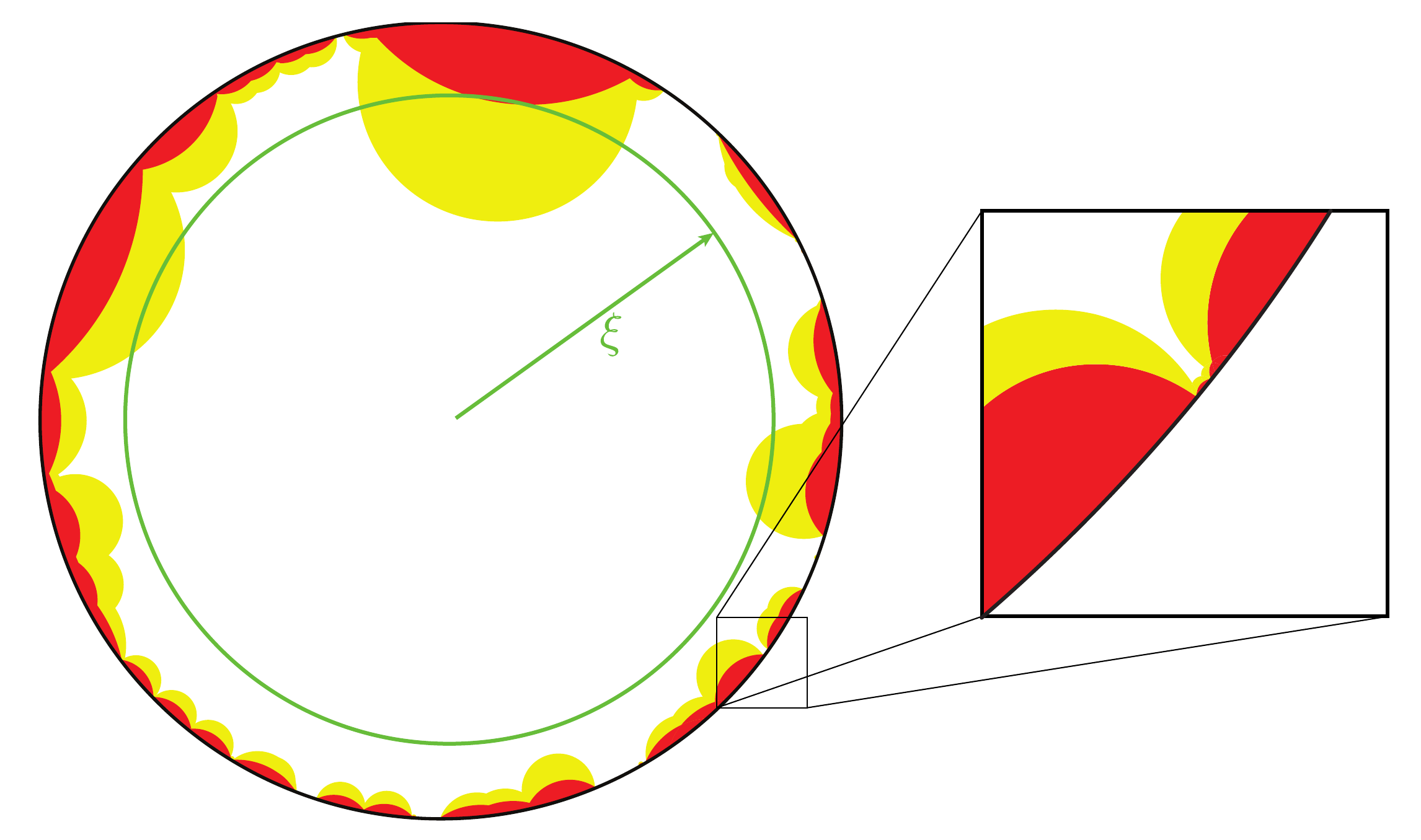}
\caption{A fixed-time slice of our universe in a compact (Poincar\'{e}) representation.  Collisions with an infinite number of other bubbles create a fractal web of yellow (light gray) and red (dark gray) regions that surrounds the exterior of our bubble.  Because of the compact representation, most of the volume lies around the rim of the disk where the structure is fractal, as can be seen in the inset.  To compute the ratio of white to yellow, the plan of attack is to first perform an angular average to determine the fraction of the surface of a `test sphere' of radius $\xi$ that is each color; the radial average is carried our subsequently.} 
\label{FracUni}
\end{center}
\end{figure}

However, there will not just be one collision.  Instead, because each bubble is surrounded by an infinite volume sheath in which other bubbles can form, there will be an infinite number of collisions and the entire exterior of our bubble's light-cone will be peppered with observable scars.  The end result will be a fractal web~\cite{Guth:1982pn, Winitzki:2001np} of white, yellow and red, as in Fig.~\ref{FracUni}.

\subsection{Persistence of Memory}

Fig.~\ref{FracUni} is surprising in that, although the interior of an individual bubble is homogeneous, isotropic and boost invariant, those cosmological symmetries are broken by bubble collisions.  Namely, observers closer to the rim of the disk are more likely to be yellow or red, despite the fact that this concept is not boost invariant.  This is a consequence of the ``persistence of memory'' effect, identified in GGV~\cite{Garriga:2006hw}.  The effect traces back to the fact that the universe (or any system) cannot remain remain trapped in a false phase forever into the past~\cite{Borde:1993xh}.  This means there must be an initial space-like slice that defines the beginning of the false phase, and its presence explicitly breaks the boost invariance of the system.

\begin{figure}
\begin{center}
\includegraphics[width=130mm]{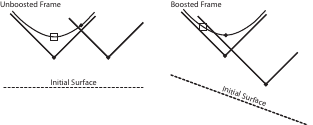}
\caption{Lorentz invariance is broken by the initial conditions slice, which defines a rest frame for the bubble.  Only one observer, represented by the square, is in the center of the bubble in this frame.  Other observers, represented by the dot, are far from the center and are more likely to see a collision.  A boost that translates such an observer to the center of the bubble also affects the initial conditions slice, which resolves any possible paradox.}
\label{PerstMem}
\end{center}
\end{figure}

This situation is easy to understand in Minkowski space (with gravity turned off).  Imagine preparing a metastable state in which the field lies in the false vacuum everywhere at some time $t_0$, and then allowing bubbles to nucleate, expand, and collide.  Lorentz invariance is broken by the fact that there is an initial time slice at $t_0$ on which there are no bubbles.  This surface defines a natural `center' to any bubble; henceforth, the term center is used to refer to the point in the bubble that is at rest with respect to the initial time slice (represented by the square in Fig.~\ref{PerstMem}).  The broken symmetry manifests itself in the fact that observers who are far from the center are more likely to see a collision than observers near the center of the bubble.  This can be seen easily in Fig.~\ref{PerstMem}; a boost that translates such an observer along the hyperboloid so that they seem to occupy the center also affects the initial condition slice.  Therefore, the observer who is far from the center expects to see more bubbles and larger bubbles from one side.  For this observer, the anisotropy is directly linked to the initial surface, and memory of the initial condition lasts forever in this way.

When gravity is turned off, the initial condition slice is necessary because the phase transition completes after a finite amount of time.  With eternal inflation, that is no longer the case (which is why it is called eternal).   However, it has been shown that eternal inflation can only be eternal into the future, since it is always past geodesically incomplete~\cite{Borde:2001nh}.  In other words, an initial condition slice is still necessary.  Although formally the initial surface can be sent to $t_0 \rightarrow - \infty$, it still cuts off half of the de Sitter hyperboloid and breaks the boost part of the symmetry group.  So the same conclusion emerges: observers near the wall are more likely to see bubble collisions than those near the center, as can be seen in Fig.~\ref{FracUni}.  This means that, if he sees enough collisions, an observer can determine his location inside the bubble.  Though just seeing one collision is enough to confirm eternal inflation, the more bubbles he sees, the more he learns about the initial conditions.  For instance, if he sees more bubbles to his left than his right, he knows the center of the bubble lies to his right.

\subsection{Plan of attack}

The quantity of interest is the ratio of the white volume to yellow volume, ignoring red volume, where white refers to the volume containing observers who are causally disconnected from all collisions; yellow, to those who see only collisions with angular scale less than $\pi$; and red, to those who see angular scale greater than $\pi$.  The plan of attack is as follows: for a given time-slice of our bubble, like the one drawn in Fig.~\ref{FracUni}, each bubble collision `paints' sections of this slice either yellow or red.  After an infinite number of collisions, the slice will be painted with a fractal web of white, yellow, and red volumes.   Consider a test sphere of radius $\xi$ about the natural center of that slice.  Its surface will have splotches of each of the three colors due to collisions that converted regions of white at that radius to yellow or red.  In section 3, the fraction of the sphere's surface that is each color is determined as a function of the radius of the test sphere.  This is a well-defined quantity, which represents an average over the angular direction.  The calculation is unambiguous, because all quantities are finite. In section 4, the average of test sphere results is taken over the radial direction $\xi$.  At this step, divergent volumes are encountered, and a discussion of the measure problem is taken up.

\section{Angular average over bubble collisions}

To compute the angular average over bubble collisions for a fixed distance $\xi$ from the center of the bubble, the same coordinate notation as AJS~\cite{Aguirre:2007an} is used.  Their setup is reviewed below.

\subsection{Coordinate system}

The exterior false vacuum space is de Sitter, which can be obtained by embedding a hyperboloid in one extra dimension. Taking coordinates $X_\mu$, $\mu=0,...,4$, and Minkowski metric $ds^2=\eta_{\mu \nu} X^\mu X^\nu$, for the embedding space, then the desired metric is induced on the surface $\eta_{\mu \nu} X^\mu X^\nu =H_{\rm{F}}^{-2}$.  A standard set of coordinates, typically referred to as the `flat slicing' of the hyperboloid, are obtained by the transform 
\begin{eqnarray}
X_0&=& H_{\rm{F}}^{-1}\sinh H_{\rm{F}} t + \frac 1 2 H_{\rm{F}} 
e^{H_{\rm{F}} t}r^2 \nonumber \\
X_i &=& r e^{H_{\rm{F}} t} \omega_i \nonumber \\
X_4 &=& H_{\rm{F}}^{-1} \cosh H_{\rm{F}} t - \frac 1 2 H_{\rm{F}} 
e^{H_{\rm{F}} t}r^2,
\end{eqnarray}
where 
$(\omega_1,\omega_2,\omega_3)=(\cos\theta,\sin\theta\cos\phi,\sin\theta\sin\phi)$.  
Under these coordinates, the metric takes the familiar form
\begin{equation}
ds^2=-dt^2+e^{2H_{\rm{F}} t}[dr^2 + r^2 d\Omega_2^2].
\end{equation}

Another useful coordinate system, and the one in which most of the calculations are performed, is the conformally compact one used to draw a Penrose diagram.  These coordinates are also useful because they cover the entire hyperboloid in embedding space, whereas the `flat slicing' only covers half.  The coordinates $(T,\eta,\theta,\phi)$ are given by
\begin{eqnarray}
X_0&=& H_{\rm{F}}^{-1}\tan T \nonumber \\
X_i &=& H_{\rm{F}}^{-1} \frac{\sin\eta}{\cos T} \omega_i \nonumber \\
X_4 &=& H_{\rm{F}}^{-1}\frac{\cos\eta}{\cos T} ,
\end{eqnarray}
where $-\pi/2\le T\le\pi/2$ and $0\le\eta\le\pi$.  The de Sitter 
metric in these coordinates is
\begin{equation}
ds^2=\frac{1}{H_{\rm{F}}^2 \cos^2T}[-dT^2+d\eta^2+\sin^2\eta \, 
d\Omega_2^2].
\end{equation}
This measure diverges when $T\rightarrow\pm \pi/2$, because an infinite expanse in time has been condensed to a compact coodinate.

As noted above, because eternal inflation is geodesically incomplete, there must be an initial value surface at some time $t$ in the past on which no bubbles were present, where $t$ here is the time coordinate of the flat slicing.  Pushing this surface all the way back to $t\rightarrow - \infty$ corresponds to
\begin{equation}
T=\eta-\pi/2.
\end{equation}

\begin{figure}
\begin{center}
\includegraphics[width=90mm]{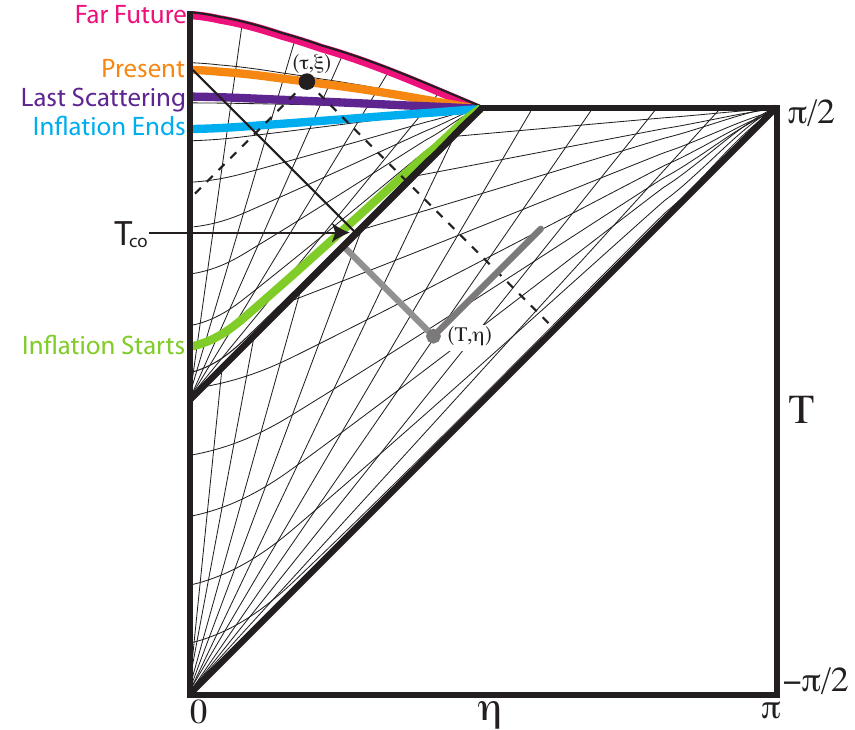}
\caption{A Penrose diagram representing one bubble embedded in an exterior de Sitter space.  An observer at position $(\tau,\xi)$ inside the bubble looks back and sees a collision from a bubble that nucleated at $(T,\eta)$ in the exterior space.  Several time slices inside the bubble are shown in various colors.  Each time slice inside the bubble can be identified by a parameter $T_{\rm{co}}$, which represents the coordinate $T$ at which the backwards light-cone of the central observer intersects the bubble wall, illustrated here for the slice marked ``Present''.  The ``Inflation Ends'' slice is drawn sagging slightly below horizontal because in this figure $H_\text{I}$ is assumed to be roughly $H_\text{O}$.  If $H_\text{I}$ were much smaller, this slice would instead bulge way up into the hat; and, likewise, the smaller $\Lambda$ is in comparison to $H_\text{O}$, the pointier the hat.}
\label{CoordFig}
\end{center}
\end{figure} 

A bubble can be included in this space-time, like in Fig.~\ref{CoordFig}.  Coleman and De Luccia~\cite{Coleman:1980aw} gave a prescription for finding the exact form of the post-nucleation bubble interior by analytically continuing a spherically symmetric instanton.  In their analysis, the null cone corresponds to the field value on the other side of the barrier.  Inside the light cone, the metric is that of an open FRW cosmology, with metric
\begin{equation}
ds^2=-d\tau^2+a^2(\tau)[d\xi^2+\sinh^2\xi d\Omega_2^2],
\end{equation}
where $a(\tau)$ is just the scale factor which time evolves by the Friedmann equations.  This metric is induced by the embedding
\begin{eqnarray}
X_0&=& a(\tau) \cosh \xi  \nonumber \\
X_i &=& a(\tau) \sinh\xi \omega_i \nonumber \\
X_4 &=& f(\tau),
\end{eqnarray}
where $f(\tau)$ solves the differential equation $f'(\tau)^2+1=a'(\tau)^2$.  If it is an empty bubble with only true vacuum, then $a(\tau)=H_{\rm{T}}^{-1}\sinh(H_{\rm{T}}\tau)$ and $f(\tau)=H_{\rm{T}}^{-1}\cosh(H_{\rm{T}}\tau)$.  This gives the `open slicing' of a de Sitter hyperboloid.  This hyperboloid, with curvature $H_{\rm{T}}$ must be pasted into the false vacuum hyperboloid, with curvature $H_{\rm{F}}$, along the bubble wall.  To do this, one has to choose a direction along which to paste, which breaks the original SO(4,1) symmetry of de Sitter space down to SO(3,1).

In Fig.~\ref{CoordFig}, our bubble is defined to be at $(t=0,r=0)$ in the false vacuum space, or $(T=0, \eta=0)$.  Inside the bubble, the coordinates are $(\tau, \xi)$ as well as two angular variables.  Equal $\tau$ slices can be identified by the parameter $T_{\rm{co}}$, which refers to the value of the $T$ coordinate at which the backwards light cone of the center observer intersects the bubble wall.  The parameter $T_{\rm{co}}$ lies between the values $0$, at the big bang, and $\pi/2$, which is only reached in the far future of asymptotically Minkowski universes.  Until this point, the entire set-up is identical to AJS~\cite{Aguirre:2007an}.

A useful coordinate substitution for calculations is 
\begin{eqnarray}
\label{sub}
u&=&\tan \left(\frac{\eta-T}{2}\right) \nonumber \\
v&=&\tan \left(\frac{\eta+T}{2}\right).
\end{eqnarray}
Under this transform, the integration measure becomes 
\begin{equation}
\label{measure}
\frac{\sin^2\eta}{\cos^4 T}\sin\theta\,\, 
\rm{d}\eta\,\rm{d}T\,\rm{d}\theta\,\rm{d}\phi = 2 
\frac{(u+v)^2}{(1+uv)^4} \sin\theta\,\, 
\rm{d}u\,\rm{d}v\,\rm{d}\theta\,\rm{d}\phi.
\end{equation}

\subsection{Angular average}

\begin{figure}
\begin{center}
\includegraphics[width=80 mm]{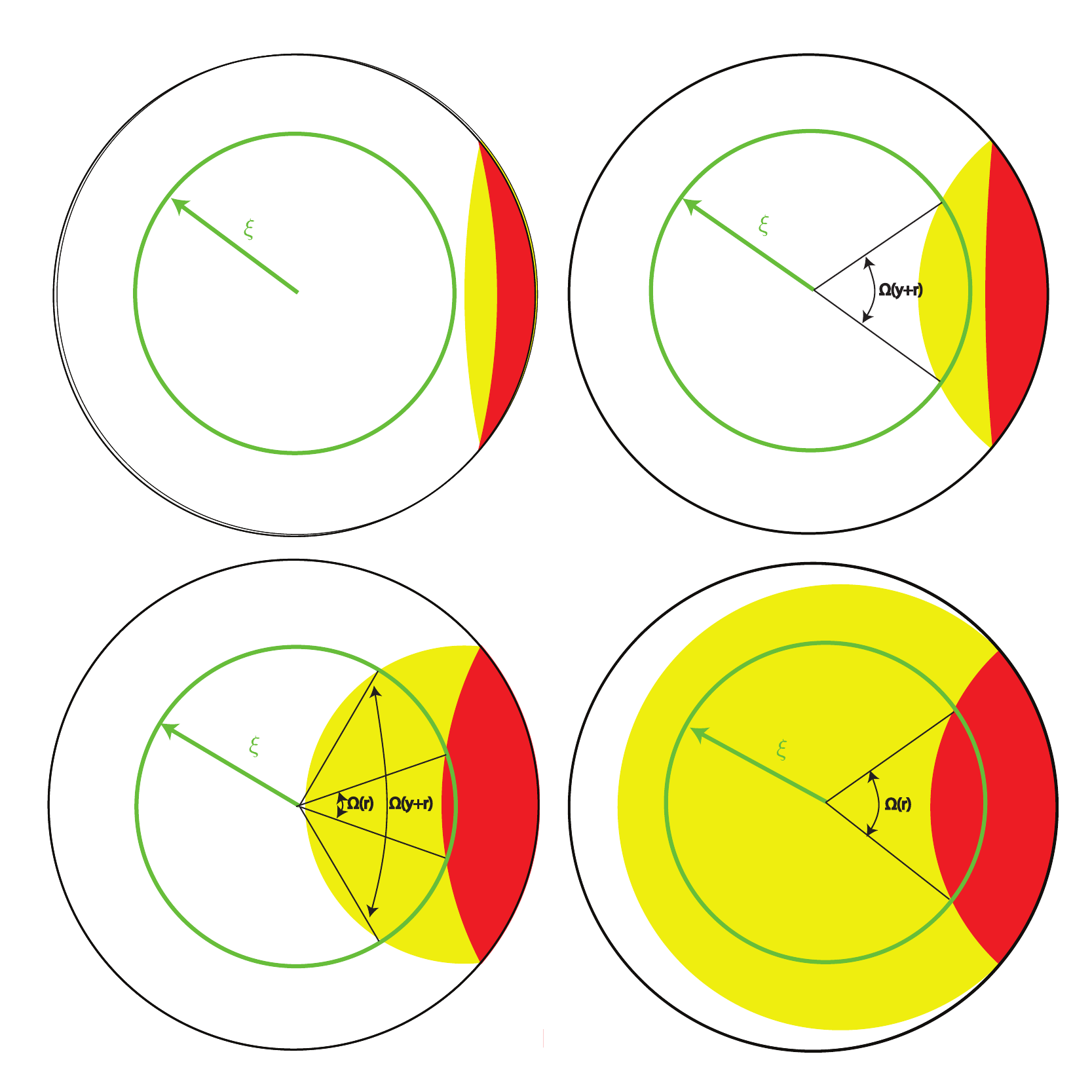}
\caption{Consider a `test sphere' or radius $\xi$ drawn about the center of the bubble.  As time passes, the yellow and red regions from a single collision encroach into the bubble and increasingly overlap, or `paint', the surface of our test sphere.    A bubble that collides with ours will `paint' the surface of our test sphere of radius $\xi$ either yellow (light gray) or red (dark gray).  This defines the quantity $\Omega(\xi,u,v)$.}
\label{OneColl}
\end{center}
\end{figure}

Consider a sphere of radius $\xi$ drawn around the center of our bubble.  Each collision will `paint' the surface of the sphere with yellow and red splotches.  We are interested in the expected fraction of the sphere that is painted by each color and the first step will be to consider a single collision.  Fig.~\ref{OneColl} shows how one collision affects the test sphere at various values of $T_{\rm{co}}$.  With time, information about the collision propagates inward and the yellow and red regions are seen to expand.  When $T_{\rm{co}}$ is small enough, the sphere is unaffected.  As $T_{\rm{co}}$ increases, an amount of solid angle $\Omega$ gets painted and eventually, if $T_{\rm{co}}$ is allowed to get big enough, the whole sphere will be painted and $\Omega$ will equal $4\pi$.  There are two quantities we will want to compute: $\Omega(\rm{y+r})$, which represents the amount of solid angle that is painted either yellow or red, and $\Omega(\rm{r})$, which the amount of solid angle that is just painted red.

Fortunately, most of the hard work to obtain these quantities was done by AJS~\cite{Aguirre:2007an}.  Imagine an observer at $(\tau, \xi, 0,0)$ who is observing a collision that occurred at $(u,v, \theta, \phi)$.  AJS calculated the angular scale $\Psi$ of the collision that he would measure (their equation 23).  If you set $\Psi=0$ and solve for $\theta$ in that equation, you find the largest possible angle $\theta$ that an observer can be away from the collision and still be in its light cone.  Then, $\Omega(\rm{y+r})= 2 \pi(1-\cos(\theta))$.  Likewise, setting $\Psi=\pi$ lets you solve for $\Omega(\rm{r})$.  The results are given by:
\begin{eqnarray}
\Omega(\xi,u,v)(\rm{y+r})&=&2\pi\left(1-\frac{v-u}{v+u}\coth \xi + 
\frac{\Ts-uv}{\T (v+u)\sinh\xi}\right) \nonumber \\
\label{omegaxi}
\Omega(\xi,u,v)(\rm{r})&=&2\pi\left(1-\frac{v-u}{v+u}\coth \xi + 
\frac{-2uv}{\T (v+u)\sinh\xi}\right),
\end{eqnarray}
provided these functions are both positive and smaller than $4\pi$.  When the above functions give negative values, those values should be replaced by $0$; and likewise values greater than $4\pi$ should be replaced by $4\pi$.

\begin{figure}
\begin{center}
\includegraphics[width=80 mm]{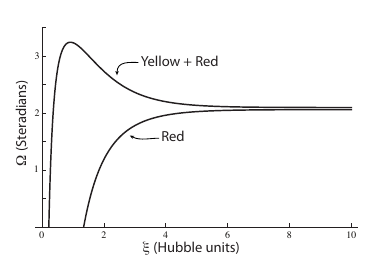}
\caption{As its radius $\xi$ is increased, the test sphere increasingly overlaps with the yellow or red region, and is thus `painted' either yellow or red. The number of steradians of solid angle $\Omega$ that have been painted is plotted against the sphere radius $\xi$ (in Hubble units).  In this case, the collision bubble is taken to have nucleated at $(u,v)=(.25, 1.25)$ and $T_{\rm{co}}=\pi/4$.}
\label{OmegaPlot}
\end{center}
\end{figure}

These results are plotted as a function of $\xi$ in Fig.~\ref{OmegaPlot} for a sample value of $u$, $v$, and $T_{\rm{co}}$.  At small values of $\xi$, the sphere does not intersect the yellow or red region, so $\Omega=0$.  As $\xi$ is increased, it eventually intersects the yellow region, then the red region.  The area of intersection grows and eventually asymptotes a fixed value.  In some cases, the colored region can completely envelop the test sphere, giving the value $\Omega=4\pi$.

It turns out to be computationally easier to ignore the fact that two collisions can create overlapping yellow patches, and then to subsequently correct for the mistake.  Ignoring the overlaps results in over-counting, and so the answer turns out to be greater than $4\pi$.  At the same time, the fact that red paint covers yellow paint in any regions where the overlap is ignored.  Both effects are taken into account at a later step in the computation.  

The expected total solid angle of a given color, with double-counting of overlaps, is given by the integral
\begin{equation}
\Omega_{\rm{tot}} (\xi)= 4 \pi \int \hspace{-2.5mm} \int 
\Omega(\xi,u,v) \times P(u,v) \,du\, dv,
\label{omegatot}
\end{equation}
where $\Omega$ is capped below by $0$ and above by $4\pi$, as described above.  $P(u,v)$ is the probability that a bubble nucleated at a position $(u,v)$, which is just given by the nucleation rate per spacetime volume $\Gamma$ times the metric factor in Eq.~\eqref{measure}.  This integral is the contribution from a bubble that nucleated at position $(u,v)$ times the probability that it nucleated there.

\begin{figure}
\begin{center}
\includegraphics[width=\textwidth]{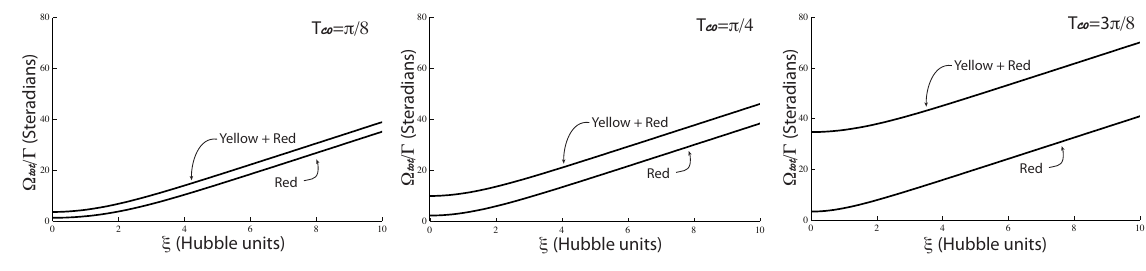}
\caption{An infinite number of collisions are considered and weighted by their liklihood.  The total number of steradians of solid angle that are painted each color, $\Omega_{\rm{tot}}$, is plotted as a function of $\xi$ (in Hubble units) for various values of $T_{\rm{co}}$.  Overlaps are explicitly being ignored, which is why both functions diverge at large $\xi$ instead of asymptoting $4\pi$.  The essential feature is that the plots quickly become linear with the same slope.  The only attribute that $T_{\rm{co}}$ appreciably affects is the distance between the two curves.}
\label{OmegaTotPlot}
\end{center}
\end{figure}

The integrals evaluate to
\begin{align}
\label{omega}
\Omega_{\rm{tot}}(\rm{y+r}) &= \frac{(4\pi)^2 \Gamma}3 \left(\Ts + 
\ln(1+2 \T \cosh \xi + \Ts) \right)  \nonumber  \\
\Omega_{\rm{tot}}(\rm{r}) &= \frac{(4\pi)^2 \Gamma}3 \bigg[ 
\frac{\Ts+(2-3\Ts)B(\T)}{2(\Ts-1)} \\[5pt]
{}&\qquad \qquad \qquad \qquad \qquad + \ln\left(\frac{2+2 \T 
\cosh\xi}{\T}\right)\bigg],  \nonumber 
\end{align}
where $B(x)$ is a function over all positive $x$ that is everywhere real, given by
\begin{equation}
B(x)=\frac{1}{\sqrt{x^2-1}}\sec^{-1} x.
\end{equation}
These two functions are plotted in Fig.~\ref{OmegaTotPlot}.  For all values of $T_{\rm{co}}$, the behavior is the same: they start at some positive value and quickly approach a line of constant slope given by $(4\pi)^2\Gamma/3$; importantly, the difference between the two lines approaches a constant.  This function diverges at large $\xi$ because of double-counting of overlaps.  

Now, let's discuss how to account for overlaps.  First, consider the case where there is only one color of paint, and also a lot of it, far more than is necessary to paint the surface of the sphere (as is the case at large $\xi$).  The paint is randomly dribbled over the surface of the sphere.  The more paint that is already down, the more likely it is that there will be an overlap.  For this reason, the expected fraction of unpainted sphere is exponentially small in the amount of paint.  In our case, there are actually two colors of paint, but they can be treated separately.  The probability of a point on the sphere being unpainted, and therefore white, $P(\rm{white})$, is exponentially small in the total amount of paint; likewise, $P(\rm{white})+P(\rm{yellow})$  is exponentially small in the amount of red paint.  This gives rise to the following equations:

\begin{eqnarray}
\label{prob}
P(\rm{white})&=&e^{-\Omega_{\rm{tot}}(\rm{y+r)}/4\pi}, \nonumber \\
P(\rm{yellow})&=&e^{-\Omega_{\rm{tot}}(\rm{r)}/4\pi}-e^{-\Omega_{\rm{tot}}(\rm{y+r)}/4\pi}, \nonumber \\
P(\rm{red}) &=& 1-e^{-\Omega_{\rm{tot}}(\rm{r)}/4\pi},
\end{eqnarray}
where the $\Omega_{\rm{tot}}$'s are given in Eq.~\eqref{omega}.  

\section{The odds of being yellow (radial average)}
\begin{figure}
\begin{center}
\includegraphics[width=70mm]{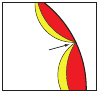}
\caption{Despite the fact that the rim of the Poincar\'{e} disk is predominantly red, a few points of measure zero are not, which can occur when a finger of white stretches all the way out to the bubble wall.  Because of the diverging volume element at the rim of the disk, an infinite volume of white and yellow is actually contained in the tip and so it is overwhelmingly likely that the Earth lies in one.}
\label{Finger}
\end{center}
\end{figure}

In the previous section, the probability that a point a distance $\xi$ from the center of our bubble will be either white, yellow, or red was computed.  Both the probability of being white and the probability of being yellow were shown to taper off with the same rate, $\exp(-4 \pi \Gamma \xi/3)$. However, the volume element grows as $\sinh^2 \xi$, which dominates this exponential decay.  So, despite the fact that the probability of being either white or yellow approaches zero at large $\xi$ (compared to the probability of being red), the volume of both colors diverges because there remain a few points of measure zero on the wall that are uncolored.   `Fingers' of white and yellow reach out and touch these points, as in Fig.~\ref{Finger}, and an infinite volume is contained in these exponentially thin fingers.  Because of the divergent volume, the overwhelming weight of probability comes from these fingers, and all other structure becomes irrelevant.

As was argued above, the relevant quantity is the ratio of yellow volume to white volume, where red volume is ignored. Both volumes are infinite, so a ratio needs to be formed, and the presence of the fingers has suggested how it should be taken.  Since the divergence in the volume lies at the tip of the finger, it suggests the definition
\begin{equation}
\label{f}
f\equiv \lim_{\xi \to\infty} \frac{P(\rm{yellow})}{P(\rm{white})+P(\rm{yellow})}= 1-\lim_{\xi \to\infty} e^{-(\Omega_{\rm{tot}}(\rm{y+r})-\Omega_{\rm{tot}}(\rm{r}))}.
\end{equation}
The contributions to $f$ from finite $\xi$ have become unimportant, which amounts to comparing the coefficients in front of the divergent parts of the two volumes.  This followed from the fact that the Earth is overwhelmingly likely to be located in a finger structure like the one shown in Fig.~\ref{Finger}, where the fractal touches the rim of the disk; this seems like the only natural definition for $f$.  

It is important to mention that this involves making a choice for the volume measure on the bubble.  Other measures have been suggested (the causal patch measure, the co-moving probability measure, and the scale-factor measure, etc.) which weight the inside of the bubble differently.  Most examples tend to favor the center of the bubble over the exterior, and more will be said about them in the following section.

Substituting Eq.~\eqref{prob} and Eq.~\eqref{omega} into Eq.~\eqref{f} gives
\begin{equation}
\label{fapprox}
f=1-(\T)^{-4\pi\Gamma/3} \times \exp\left[-4 \pi \Gamma/3\left(\Ts-\frac{\Ts+(2-3\Ts)B(\T)}{2\Ts-2}\right)\right].
\end{equation}
A `reasonable' universe, will have an exponentially small value for $\Gamma$, which allows us to Taylor expand.  
\begin{equation}
\label{fmoreapprox}
f\approx\frac{4\pi \Gamma}{3}\left(\log\T +\Ts-\frac{\Ts+(2-3\Ts)B(\T)}{2\Ts-2}\right)
\end{equation}
Because $f$ is proportional to $\Gamma$, which is exponentially small, $f$ will be small unless the quantity in parentheses becomes large.  $T_{\rm{co}}$ has to be between $0$ and $\pi/2$ (and the point at exactly $T_{\rm{co }}=0$ must be excluded because all points are at $\xi=0$).  Since cosmic evolution inside our bubble is dominated by inflation, we can replace $T_{\rm{co}}$ by the value it has in a pure de Sitter universe, which is $\arctan(H_{\rm{O}}/H_{\rm{I}})$, where $H_{\rm{O}}$ is the scale of inflation outside the bubble and $H_{\rm{I}}$ is the Hubble scale for the second round of inflation that occurs inside the bubble.  In this case, saving only the largest term, 
\begin{equation}
f\approx\frac{4\pi \Gamma}{3} \left(\frac{H_{\rm{O}}}{H_{\rm{I}}}\right)^2.
\end{equation}
As stated in the introduction, this result is very similar to the one found by Freigovel \emph{et al.}~\cite{Freivogel:2009it} for collisions with bubbles that had cosmological constants that were equal to the the exterior value.  Presumably, this means that collisions with any type of bubble will be proportional to the same factor, with a numerical constant that depends on the difference in cosmological constant between the two bubbles.

\section{Discussion and conclusions}
In the previous section, it was shown that the probability that the Earth is in the causal future of a collision is proportional to $\Gamma (H_{\rm{O}}/H_{\rm{I}})^2$.  Could this factor ever be big enough that we might observe one?  The factor in parentheses is roughly $(10^{18} \text{ GeV}/10^{15} \text{ GeV})^2\sim 10^{6}$, but could be bigger with further fine-tuning in the slow roll parameters.  The nucleation rate will generically be smaller than this, and bubble collisions will not be observable.  However, it is possible to imagine that, out of the many possible tunneling directions in the landscape, there could be a handful with decay rates that are big enough for this to be sizeable.  In that case, the Taylor expansion used in Eq.~\eqref{fmoreapprox} would not be valid, and $\Gamma (H_{\rm{O}}/H_{\rm{I}})^2$ would be proportional to the number of bubble collisions in our past light-cone.  Since the nucleation rate depends exponentially on the Euclidean action of the bounce solution, it requires fine-tuning for that number to be order one.  Instead, the typical thing to expect is to be in the future light-cone of a large number of collisions from the same few bubble types and none of any other type.

However, even if an observer meets this condition for observation, it does not guarantee that he can detect any evidence of a collision.  Instead, yellow observers can be split into three categories.  The first see an unacceptably large level of anisotropy.  The isotropy of the CMB guarantees that Earth does not lie in this region; in fact, if the probability that Earth lies in this region were large, eternal inflation would be ruled out.  A second category includes observers for whom the signal of the bubble collisions has been effectively erased, beyond any conceivably detectable level.  In order for observations to confirm eternal inflation and the multiverse, the Earth must lie in the third region, the in-between one, from which collisions are observable, but just beneath the level of current detection.  

A good understanding of the consequence of a bubble collision is required to evaluate the probability that Earth is in this third region.  Although this understanding does not exist, two points are clear.  First, the requirement that we live in a finger constrains the possible observational signatures.  Second, current understanding is sufficient to show that the odds of observing the multiverse depend sensitively on the details of the shape of the energy landscape.

\subsection{The view from inside a finger}

The requirement that we live in a finger already constrains the possible observational signatures.  For instance, from the vantage point of a finger, only certain types of collisions will be visible.  In particular, being in a finger eliminates the possibility of observing the simultaneous collision of more than three bubbles, because if there are too many overlapping bubbles then there is no finger.  Multiple bubble collisions were studied as a way to produce black holes~\cite{Moss:1994pi}, but such an effect will not be visible to us inside a finger.  

Also, since the argument in the previous section relied so heavily on the presence of a finger, it is worth discussing the case when the fingers are not present.  Recently, several measures have been suggested~\cite{DeSimone:2008if, Bousso:2006ev, Linde:2006nw, Susskind} which, in order to regulate the infinities of eternal inflation, have different volume-weightings within the bubble interior.  In particular, these measures tend to introduce a cut-off value of $\xi$ that chops off the fingers, or that suppresses their contribution to the total volume, so that the total volume on the interior of the bubble is finite.  If, indeed, something cut off the fingers at a radius $\xi_{\rm{max}}$, the answer becomes cutoff-dependent and can be computed directly by not taking the $\xi\rightarrow\infty$ limit.

\subsection{Observability}

In order to be observable, the collision must either leave a remnant that survives the round of inflation that occurs within the bubble, or it must alter the way in which inflation proceeds.  Radiation or particles created at the collision will, in general, not be observable, because they will be exponentially dilated by the round of inflation inside the bubble.  Instead, the most likely observational effect will involve a local change in the end-time of inflation, or a perturbation to the shape of the reheating surface, which would look like a hot- or cold-spot in the CMB.

These thoughts were recently made more concrete by explicit analytic and numerical analysis in a particular model of a bubble collision~\cite{Freivogel:2007fx, Chang:2007eq, Chang:2008gj, Aguirre:2007wm, Aguirre:2008wy}.  In this model, the two bubbles stick together and an infinitely thin cone of radiation, like a shock-wave, carries away the remaining energy and momentum of the colliding walls.  While the validity of this assumption is not established, it allows for direct computation, and it has proven very useful for identifying the important effects.  In this model, the case of collisions with identical bubbles, was studied numerically~\cite{Aguirre:2008wy} and it was shown that inflation within the yellow region and the shape of the re-heating surface are nearly unaffected by the collision.   This suggests that even if $f$ is large, the odds of observing such collisions are essentially zero---the collisions are not sufficiently observable.  However, the jury is still out on whether other gravitational effects, or a more accurate coupling of the inflaton field to the radiation emitted at the collision might lead to new observable signatures.

In summary, it is no more likely to have a large number of same-bubble collisions in our past light-cone than any other variety; it all depends on the nucleation rate for the bubble.  (The fact that we live in the parent vacuum that we do, and that our bubble has such a low cosmological constant are two arguments for why the nucleation rate for our bubble might be high, but neither is that convincing.)  Current understanding of bubble collisions suggests that same-bubble collisions will be harder to detect, but this is still not conclusive.  The next thing to consider is a wider spectrum of observational signatures.  For one thing, the model considered above ignores the possibility of the walls passing through each other and oscillating, as in~\cite{Hawking:1982ga}.  But more to the point, the examples considered so far are only the tip of the iceberg.   In a true landscape, there is a smorgasbord  of possible observable remnants of collisions.  There will be collisions with bubbles where the constants of nature take different values, which would create waves of death, across which the laws of physics change.  Collisions could cause extra dimensions to decompactify.  Even the value of Newton's constant could change between bubbles, so that bubbles with the same value of the cosmological constant will inflate differently.  Collisions could produce domain walls that themselves inflate, producing huge volumes of anisotropic space.   Higher dimensional effects could also become important, or quantum effects like resonant tunneling.  Such exotic possibilities could produce unacceptable levels of anisotropy, and pose a threat to inflationary theory.  The fact that they have not been detected shows that  $\Gamma \ll (H_{\rm{O}}/H_{\rm{I}})^2$ for such dangerous bubbles.

\section*{Acknowledgments}
It is a pleasure to thank Paul Steinhardt for his help at all stages in this project.   I would also like to thank Leonard Susskind, Matt Johnson, Tiberiu Tesileanu, and Anthony Aguirre  for useful discussions, and Elizabeth Dahlen for comments on the manuscript.  This work is supported in part by the US Department of Energy grant DE-FG02-91ER40671.

\bibliographystyle{unsrt}
\bibliography{mybib}

\end{document}